\documentclass[final]{aipproc}
\layoutstyle{6x9}

\newcommand{\azero}{\mbox{$A_{0}$}}
\newcommand{\tanb} {\mbox{$\tan\beta$}}

\begin{document}

\begin{flushright}
ACT-02/09\\
MIFP-09/05\\
\end{flushright}

\title{The MAGIC of SSC and how it affects LHC}


\author{Nick E. Mavromatos}{
  address={King's College London, Department of Physics, Strand, London WC2R 2LS, U.K.}
}

\author{D.V. Nanopoulos}{
  address={Department of Physics,
Texas A \& M University, College Station,
TX~77843-4242, U.S.A. \\
  Astroparticle Physics Group, Houston
Advanced Research Center (HARC), Mitchell Campus,
Woodlands, TX 77381, U.S.A. and \\
Academy of Athens,
Chair of Theoretical Physics,
Division of Natural Sciences, 28~Panepistimiou Avenue,
Athens 10679, Greece.}
}

\begin{abstract}
We discuss the phenomenology of Supercritical String Cosmology (SSC) in the context of Dark Matter constraints on
supersymmetric particle physics models at LHC. We also link our results with recent findings of the MAGIC,
H.E.S.S. and Fermi Telescopes on delayed arrival of highly energetic photons from the distant Galaxies and GRBs. The link is provided by a concrete model of space-time foam in (supercritical) string theory, involving space-time defects and their interaction with matter in a brane world scenario.

\vspace{0.5cm}
\begin{center}
\emph{Expanded and updated version of talk presented by D.V.N. at Dark Matter 2008, 8th UCLA Symposium on Sources and Detection of Dark Matter and Dark Energy in the Universe,\\ Marriott Hotel, Marina del Rey, California (U.S.A.), February 20-22 2008}
\end{center}

\end{abstract}

\maketitle

%%%%%%%%%%%%%%%%%%%%%%%%%%%%%%%%%%%%%%%%%%%%
%% MAINMATTER
%%%%%%%%%%%%%%%%%%%%%%%%%%%%%%%%%%%%%%%%%%%%

{\bf Supercritical String Cosmology (SSC) Framework:} Supercritical String Cosmology (SSC) is a framework, largely phenomenological at present,
developed in \cite{ssc}, in an attempt to discuss cosmological models in non-equilibrium
string models of the Universe. Such departures from equilibrium may be attributed  to cosmically catastrophic events (Big-Bang, brane collisions in brane-world scenarios \emph{etc}.) or quantum space-time foam situations, involving singular quantum fluctuations at microscopic (string) length scales.
Formally they are described by means of non-conformal deformations on the world-sheet $\sigma$-model in a (perturbative) first-quantised framework, which describes the dynamics of string excitations of a
string Universe in a near equilibrium situation. Such conditions may characterise epochs of the universe where the non-equilibrium phenomena are relatively weak, such as late eras, long after the Big-Bang or  quantum-foam vacua, where the singular fluctuations of space time are sufficiently dilute~\footnote{In some models, such fluctuations could be represented by ``real'', as opposed to virtual, D0-brane defects crossing our brane world, as the latter moves in a higher-dimensional bulk~\cite{emnw}.}. Stronger effects are much more complicated to handle analytically, as they involve strongly coupled, non-perturbative string theory, but in principle they can be present, especially in early epochs of the string Universe.

The \emph{supecritical} nature of SSC refers to the fact~\cite{aben} that in the above situations there is a central charge \emph{surplus}, as compared to the critical-string value fixed by equilibrium situations, which the SSC asymptotes to. In this sense,  SSC describes \emph{relaxation phenomena} in the string Universe. Technically speaking~\cite{ssc}, the non-conformal world-sheet deformations necessitate the introduction of the so-called Liouville mode~\cite{ddk}. This is a fully fledged dynamical world-sheet ``\emph{time-like}'' coordinate field~\cite{aben}, whose presence is essential in restoring the lost conformal invariance. However, this is done at the cost of introducing an extra time coordinate in target space. As noted in \cite{ssc}, however, the Liouville mode may be viewed as a local renormalization group scale on the world sheet of the non-conformal strings, and as such one may choose a class of schemes in which \emph{its} \emph{world-sheet zero mode} is \emph{identified} with (a function of) the \emph{target time}. There are toy models of the string Universe~\cite{gravanis} in which such identification results in minimization of the effective potential of the low-energy field theory, and hence can be achieved dynamically.
The above-mentioned relaxation time-dependent processes thus translate to an \emph{irreversible }
\emph{renormalization-group flow} on the world-sheet of the supercritical string.

From an effective target space-time field theory point of view, SSC involves time-dependent dilaton fields, and other modifications in the so-called Einstein equations, describing the pertinent cosmology.
Near an equilibrium (``fixed'' (conformal)) point in the space of string ``theories'', the general structure of these equations reads:
\begin{equation}\label{sscbeta}
{\ddot g}^i + Q(t){\dot g}^i = -\beta^i (g^j) + \dots
\end{equation}
where the $\dots$ on the r.h.s. denote corrections that become important if one goes further away from equilibrium,
the overdot denotes derivative with respect to the ``Liouville time'', called $t$ here, which, as mentioned above is identified with (a function of) the target time~\cite{ssc},
$g^i = \{ g_{\mu\nu}(t), \phi(t), \dots\}$ are background fields, pertaining to the target-space metric, dilaton and other fields, including string matter excitations, and $\beta^i$ are the relevant renormalization-group $\beta$-functions, or the so-called Weyl anomaly coefficients,
which are proportional to \emph{off-shell} variations of a target-space effective action~\cite{ssc}.
The latter vanish at a conformal point in theory space, which thus corresponds to the usual equilibrium Einstein cosmology. The quantity $Q(t)$ is the square root of a \emph{continuously varying }, ``running'' central charge deficit, which varies along a local renormalization group trajectory in the world-sheet formalism. It vanishes at conformal points, which the SSC asymptotes to.
This quantity depends crucially on the microscopic SSC model, includes strongly-coupled string contributions from  the early Universe regime and receives contributions from
matter excitations, as a result of back reaction effects of matter fields on the space-time.
As such, it is very difficult, if not impossible, to be calculated accurately, except in certain toy models~\cite{gravanis}, and thus, for most of our analyses in phenomenologically realistic models
$Q(t)$ will be determined self-consistently by solving (often numerically and approximately) the
dynamical equations (\ref{sscbeta}).

This is an important innovation of our approach as compared to the original formulation of SSC~\cite{aben}, where the central charge took on only a discrete set of constant values and the expanding string Universe, described by SSC,  is supposed to tunnel through them by a series of phase transitions until it reaches equilibrium, and exits the expansion phase.
As mentioned previously, we have constructed toy models~\cite{gravanis}, involving brane world collisions as the main cause of departure from criticality, in which the continuous running of $Q(t)$, towards equilibrium points at which it vanishes, has been demonstrated explicitly, consistent with the restoration of the conformal invariance of the $\sigma$-model by the Liouville mode and the dynamical identification of the latter with (a function of) the target time.

{\bf SSC and its effects on LHC:} The SSC framework has important phenomenological consequences, one of the most important of which pertains to the modification of the energy budget of the Dark sector of the Universe~\cite{lmn}.
In SSC, \emph{both} the time-dependent dilatons and the extra off-shell corrections in (\ref{sscbeta}),
proportional to  $\beta^i$, affect seriously the various thermal relic abundances, through modifications of the relevant Boltzmann equations by appropriate source terms, proportional to the rate of the dilaton as well as off-shell terms proportional to $\beta^i$.
As already explained, at present the SSC models we have analysed are largely phenomenological. In particular, in \cite{lmn} we had to postulate simple equations of state for the dark matter sector, which in SSC couples non-trivially to the dilaton, in such a way that the delicate Big-Bang Nucleosynthesis (BBN) constraints at MeV energy scales are respected. This requires that, at such (brief) eras in the Universe's evolution, the central charge deficit $Q(t)$ vanishes or is diminished significantly, before rising again at later eras, a feature which can be achieved in concrete toy models~\cite{gravanis,emnw}.
Moreover, in some realistic string-theory models one may also get non-thermal dark matter contributions that could be dominant in some cases, depending on the respective parameter range~\cite{nmrev}.

\begin{figure}[ht]
\centering
\includegraphics[width=.70\textwidth, height=.3\textheight]{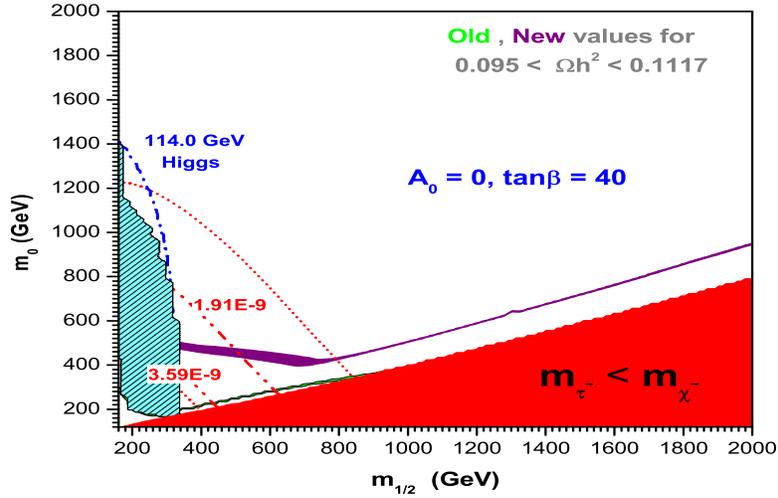}
\caption{WMAP allowed parameter space for the SSC and standard big bang cosmology shown for $\azero = 0$ $\tanb = 40$.  The very thin green (grey) band is where the neutralino relic density calculated by standard big bang cosmology agrees with the WMAP3 limits $0.0950 < \Omega  h^{2} < 0.1117$.  The thicker dark purple band shows the same agreement using the SSC calculation of the relic density.  Also shown are the Higgs mass boundary (dashed dotted blue line), muon $g_{\mu}-2$ boundaries (dashed and dotted red lines), a hatched cyan region which is excluded by $b \rightarrow s \gamma$ experimental bounds, and a lower solid red region where the neutralino is not the lightest supersymmetric particle. Picture taken from \cite{lmn,dutta}.}
\label{fig:sscpheno}
\end{figure}

Here we restrict our discussion on thermal dark matter issues. The modification of the Boltzmann equation describing the temporal evolution of the density $n$ of thermal dark matter species in SSC reads~\cite{lmn}:
\begin{equation}
{dn\over dt}+3 H n = -<\sigma v>(n^2-n_{eq}^2) + \dot\phi n + {\cal O}(\beta^i )
\label{boltz}
\end{equation}
in a standard notation, where $H$ denotes the Hubble rate of the Universe, and the suffix ``eq'' denotes
thermal equilibrium values. For weak ${\cal O}(\beta^i )$ contrbutions at late eras,
the relic density is given by~\cite{lmn}
\begin{equation}\Omega h^2=R\times (\Omega h^2)_0\end{equation}
where R is $\sim {\rm exp}\left(\int^{x_f}_{x_0}(\dot\phi H^{-1}/x )dx\right)$ and $(\Omega h^2)_0$ denotes the relic density that is obtained by ordinary cosmology. It is possible to obtain R by solving for $\phi$ the field equations for the SSC scenario. The value of $R$  is about 0.1 in order to satisfy the recent
WMAP and other observations on the evolution of dark energy in the range $0 < z < 1.6$. This new factor changes the  profile of dark matter allowed region in SUSY models and modifies significantly the relevant astrophysical constraints, as compared with the standard cosmology scenarios.  For such comparisons, one can adopt, as a simplified but concrete string-inspired low-energy  model, the minimal SUGRA (mSUGRA)
(see figure ~\ref{fig:sscpheno}). However, such studies should be viewed only as indicative, and in general one can extend the analysis to incorporate more realistic string low-energy supergravity field theories.
We investigated in detail the minimal-supergravity (mSUGRA) signals at the LHC in the
context of SSC in \cite{dutta}.
In this theory, the
presence of a time dependent dilaton  provides us with a smoothly evolving
dark energy and modifies the dark matter allowed region of the mSUGRA
model as compared to the standard cosmology. Such a dilaton dilutes  the supersymmetric
dark matter density (neutralinos) by a factor $O(10)$, as mentioned above, and consequently the
 regions with too much dark matter in the standard scenario are allowed
in the SSC. We have showed that the final states in this scenario, unlike the
standard cosmology, consist of $Z$- and Higgs-bosons and high-energy $\tau$-leptons for smaller
values of  supersymmetry (SUSY) masses allowed in this model.
We have constructed the appropriate observables and
determined the model parameters accurately. Using these parameters, we then
showed~\cite{dutta} how accurately one can determine the dark matter content and the
neutralino-proton cross section in this new scenario.
All these techniques can also be applied to determine model parameters in SSC models with different SUSY breaking scenarios, which could characterise more realistic low-energy string-inspired models~\cite{nmrev}.

{\bf MAGIC Observations:} Let us now discuss an apparently unrelated but important set of recent astrophysical observations
on the non-simultaneous arrival time of high-energy vs. low-energy photons ($\gamma$-rays) from some distant galaxies and Gamma-Ray-Bursters (GRB).
The Major Atmospheric Gamma-ray Imaging Cherenkov (M.A.G.I.C.) Telescope~\cite{magic} observed that high energy TeV photons from the active Galactic Nucleus Mkn 501, at redshift $z = 0.034$, arrive some four minutes later than their lower-energy (in the GeV energy region) counterparts.
Although the mechanism at the source for producing such high energy emission is still a mystery,
and the effect could indeed be attributed to purely conventional astrophysics, nevertheless some exotic interpretations have been proposed~\cite{magic2} as alternative explanations of the effect.
These are
based on scenarios of quantum-gravity-induced refractive indices \emph{in vacuo},
which are \emph{subluminal}, inceasing \emph{linearly} with the photon \emph{energy} and suppressed by a \emph{single power} of the of quantum-gravity energy scale.

Moreover, such explanations appeared remarkably consistent with other available data on high-energy
$\gamma$-ray emission from other more distant sources~\cite{hess,grb}, in particular, the recent observation~\cite{grb} of a 16 s delay of high-energy GeV photons from the distant (at red-shift $z = 4.35$) Gamma-Ray-Burster  GRB 080916c. The important point to notice is that in all such observations the effective quantum gravity scale is found very near the reduced Planck mass, of order $10^{18}$ GeV, which notably is the order of the string mass scale $M_s$ in traditional string theories~\footnote{However, the reader should notice that, from a modern perspective, where string and/or brane models with large extra-dimensions have been consistently constructed, the string scale is, at least at present, considered arbitrary, to be constrained phenomenologically.}.

{\bf Is MAGIC...DARK?} This latter observation motivated microscopic explanations~\cite{emn2} of these delays,
in terms of concrete models of quantum foam in the modern version
of string theory, including D-brane defects in space time~\cite{emnw}.
In particular, we have argued~\cite{emn2} that
the interaction of string matter excitations, represented as open strings attached to a three-spatial dimensional brane world, with D-particle (point-like D0-brane) defects, which cross the brane world as it propagates in a higher-dimensional bulk, cause delays which are proportional to the energy of the incident matter string,
\begin{equation}
\Delta t \propto E/M_s^2
\end{equation}
in each collision of a string state with a D-particle defect. The latter is a topologically non-trivial
process, involving temporary capture and re-emission of the matter string state.
Such delays respect \emph{causality} and are consistent with the other string space-time \emph{uncertainty relations}. Hence, they are non-perturbative effects.
Only electrically neutral excitations, such as photons, and may be neutrinos, can exhibit such non-trivial
interactions with the space-time ``medium'' of D-particles, as a result of electric charge conservation,
which remains an exact property in our models. As such, in the above framework,
only electrically neutral probes should exhibit the non-trivial optical effects of the foam `medium'.
The existence of populations of D-particle defects in a brane Universe, implies an induced refractive index of the corresponding string vacuum, and in fact there are analogies~\cite{emn2} of the situation with the standard model for a refractive index in solids, where the r\^ole of the electrons is played here by the D-particles. But there is an important difference, in that the modifications to the refractive index in the string model,  although subluminal, as in the solid-state case, are found proportional to the photon energy (frequency). The restoring force in the conventional model of the refractive index, which keeps essentially the electrons quantum-oscillating about their (average) positions, is provided in the string model by the corresponding flux strings that are stretched between the D-particle and the brane world.

Comparison with the MAGIC, H.E.S.S. and Fermi data, implies constraints on the density of the D-particle populations, which, upon assuming as a reasonable approximation that for red-shifts $z \le 5 $ the
number $n(z)$ density of the D-particle defects is red-shift independent, amount~\cite{emn2} to constraining
quantities like:
\begin{equation}
     M_s/N \ge 10^{18} ~{\rm GeV}
\end{equation}
where $N$ denotes the number of defects encountered by the string matter excitation \emph{per string length}.
The inequality is due to the uncertainties in the source mechanisms for the production of high-energy photons in the relevant measurements~\cite{magic,magic2,hess,grb}.

The presence of D-particle populations leads~\cite{emnw} also to dark energy contributions in the respective string Cosmology. It should be noted  that the interactions of string matter with the D-particles \emph{cannot} be described by equilibrium processes. The splitting and capture of the matter string by the defect, its subsequent re-emission and the associated distortion of the neighboring space-time, as a result of the D-particle recoil, can be described \emph{only} by SSC~\cite{emnw,emn2}.
In this respect, one has a microscopic model for SSC, whose parameters can then be constrained further, at least  in principle, at the LHC or other particle collider experiments in the future, such as the linear collider, as a result of the afore-mentioned induced modifications in the amount of (thermal) dark matter relics. It should be stressed at this point that, in this microscopic model of D-particle foam, the D-particle themselves serve \emph{both} as dark-energy and dark-matter objects, since they are massive objects, with masses of order $M_s/g_s$, with $g_s$ the string coupling (at late eras of the Universe), interacting only gravitationally, and thus weakly, with the string matter.
Of course, the model also has its own conventional dark matter, being a string-inspired supergravity theory. As a result, the amount of dark matter in galactic halos is also modified in this model, and such modifications can be constrained by means of gravitational lensing techniques.

\textbf{Outlook:} Although, in principle, it is possible to calculate analytically a detailed evolution of the central charge deficit of such microscopic models of D-particle foam at late eras of the Universe, it may not be a simple task to perform such calculations near the BBN epochs, where the density of D-particle defects might be much higher. In such regions, strongly-coupled SSC models are at play, for which at present we do not have
analytic control. Nevertheless, the above discussion shows how, in principle, SSC can be constrained by
diverse observations, ranging from high-energy $\gamma$-ray astrophysics to dark-matter phenomenology at particle colliders. The induced modifications in the allowed range of the parameter space for (broken) supersymmetry, but also in the Lorentz-invariance properties of the string vacuum, which
 is broken explicitly by the presence of D-particle defects on the brane world, lead to a rich phenomenology, which
could have falsifiable predictions in the near future, in a diverse range of experimental probes, both terrestrial and astrophysical.
This is the message we would like to get across with this talk, being of course completely aware of the
incomplete status and severe limitations of our SSC models at present, as far as detailed and realistic cosmologies are concerned.

%%%%%%%%%%%%%%%%%%%%%%%%%%%%%%%%%%%%%%%%%%%%
%% Sample figure:
%%
%% The option [height=...] scales the picture to the given height,
%% without it it would be printed at its nominal size
%%%%%%%%%%%%%%%%%%%%%%%%%%%%%%%%%%%%%%%%%%%%

%\begin{figure}
 % \includegraphics[height=.3\textheight]{sscpheno}
  %\caption{}
%\end{figure}

%%%%%%%%%%%%%%%%%%%%%%%%%%%%%%%%%%%%%%%%%%%%%%%%
%% BACKMATTER
%%%%%%%%%%%%%%%%%%%%%%%%%%%%%%%%%%%%%%%%%%%%%%%%

\begin{theacknowledgments}
The work of N.E.M. is partially supported by the European Union
through the Marie Curie Research and Training Network \emph{UniverseNet}
(MRTN-2006-035863) and that of D.V.N. by DOE grant DE-FG02-95ER40917.

\end{theacknowledgments}

\end{document}